\documentstyle[12pt]{article}

\newcommand{\mysection}{\setcounter{equation}{0}\section}

\begin{document}
\hfill{SMU HEP 93-08} 
\vskip 0.1cm
\hfill{ANL-HEP-CP-93-66}
\vskip 1cm
\centerline{\large\bf B-Quark Production at Hadron Colliders}
\vskip 1cm
\centerline{S. Riemersma}
\centerline{Department of Physics,}
\centerline{Southern Methodist University,}
\centerline{Fondren Science Building,}
\centerline{Dallas, TX 75275-0175, USA}
\vskip 0.3cm
\centerline{and}
\vskip 0.3cm
\centerline{Ruibin Meng}
\centerline{High Energy Physics Division,}
\centerline{Argonne National Laboratory,}
\centerline{Argonne, IL 60439}
\vskip 0.5cm
\centerline{Abstract}
\vskip 0.5cm
 
Results for b-quark production at hadron colliders, both current and
proposed, are presented.  Distributions in $p_t$ are presented for the
TeVatron and SSC.  Confirmation of agreement between the ${\cal O}(\alpha_S^3)$
calculations and UA1 data is presented, and the discrepancy between the
${\cal O}(\alpha_S^3)$ calculations and the CDF results is updated with
 the most recent data.  
\newpage 
\mysection{Introduction}
Studying B-physics at hadron accelerators requires a good understanding
of the total and differential cross sections for b-quark production.
This knowledge gives those involved in $\rm B\bar{\rm B}$ mixing, rare
B decays, and those trying to determine the CKM angles $\alpha\,, \beta
$, and $\gamma$ an idea of how many events they can expect, given the 
luminosity and the branching ratios.  It is particularly important for
those studying rare B decays as they set limits on where we can hope
to see new physics.  For these reasons and others, the complete ${\cal
O}(\alpha_S^3)$ corrections to heavy-quark production at hadron
accelerators were calculated in \cite{nde} and \cite{bkns}.  Also three groups
 \cite{ce}, \cite{lrss}, \cite{cch} have attempted to 
calculate heavy-quark production using resummation techniques in the small-$x$ 
kinematic region.  These techniques are necessary since the b-quark
mass $m_{\rm b}$ is small relative to the center-of-mass energies 
$\sqrt{S}$ of the
TeVatron and the SSC.  While these techniques offer some hope of obtaining
reasonable predictions for b-production at these machines,
the current results can best be considered as preliminary.
	
Thus we must turn to perturbative QCD for guidance, as we have no
other real choice at this point.  However, let us submit a {\em caveat} here:
fixed-order perturbative QCD works best when all the scales are
 roughly comparable, {\em i.e.}
$\sqrt{s} \approx m_{\rm b} \approx p_t$, $\sqrt{s}$ being the partonic 
center-of-mass energy. When we are not in this regime, for
example at the TeVatron and the SSC, our predictions will then be less 
reliable.
Bearing this in mind, let us continue to the results section.

\mysection{Results}

A number of fixed-target $pp$ experiments have been proposed for HERA, LHC, and
SSC.  The cross sections given in Table 1. are total cross sections
without any cuts applied.  The purpose is to give an idea of the overall
rate of b-production at these proposed experiments.  Note that these cross
sections are for inclusive b-production, so if one wants to calculate rates
for b- or $\overline{\rm b}$-production, one needs to multiply these results
by a factor of two.
\newpage
\begin{center}
Table 1. Cross Sections for Proposed Fixed-Target Experiments.
\end{center}

\begin{center}
\begin{tabular}{|l|c|c|}
\hline
$\sqrt{S}\, {\rm(GeV)}$ & Born & ${\cal O}(\alpha_S^3)$ \\
\hline
43 & 8.3 nb & 17 nb \\
124 & 0.32 $\mu$b & 0.58 $\mu$b \\
200 & 0.89 $\mu$b & 1.6 $\mu$b \\
\hline
\end{tabular}
\end{center}
These cross sections were generated using programs created by \cite{bkns} with
 the following inputs: $m_{\rm b}$ was chosen to be 4.75 GeV/
$c^2$, the mass factorization scale $M^2$ was chosen to be $m_{\rm b}^2$,
and the parton distribution set used was CTEQ1M \cite{cteq}.  We would also
like to mention here that similar results have been obtained earlier in
in \cite{bm} using a similar parton distribution set and our numbers in 
Table 1.
as well as in Table 2. below agree with theirs.  From Table 1.,
we see that the corrections even at these low energies are sizeable.  For
$\sqrt{S} = 43$ GeV, one should probably take into account resummation
effects at large-$x$ (see \cite{lsn}).
However, at these energies, we expect that the results are fairly accurate.

The situation for b-production at the TeVatron, LHC, and SSC is more
problematic.  We are no longer in a region where we expect fixed-order 
perturbative
QCD to give experimentally valid results.  Nevertheless, the predictions made
are worth noting, to get a quantitative idea of which regions in phase space
our predictions are lacking and how much of an improvement needs to be
made.  Having given sufficient warning, we present Table 2., cross sections
for the TeVatron, LHC, and SSC.
\vglue 0.25in
\centerline{Table 2. Cross Sections for the Various Colliders.}
\begin{center}
\begin{tabular}{|l|c|c|}
\hline
$\sqrt{S}\, {\rm(TeV)}$ & Born & ${\cal O}(\alpha_S^3)$ \\
\hline
1.8 & 17 $\mu$b & 37 $\mu$b \\
15.4 & 92 $\mu$b & 270 $\mu$b \\
40 & 170 $\mu$b & 550 $\mu$b \\
\hline
\end{tabular}
\end{center}
As in Table 1., no cuts were applied and the input parameters chosen were the
same.  We see rather large increases when the ${\cal O}(\alpha_S^3)$ 
corrections are
included.  The 'K-factors' are 2.2, 2.9, and 3.2 for the TeVatron, LHC,
and SSC, respectively.  The size of these 'K-factors' might give one cause
to worry, however they are slightly misleading since the massless $t$-channel
exchanges present in the ${\cal O}(\alpha_S^3)$ corrections are absent in
the Born approximation calculation.  A better indication of the convergence
should be found in comparing the ${\cal O}(\alpha_S^4)$ results with the 
${\cal O}(\alpha_S^3)$ corrections.
We were also presented with a list of cuts from various experimental groups,
and
what was settled upon was the following:  for CDF, we were asked for 
pseudorapidities $|\eta| < 1$ and $p_t > 4$ GeV/$c$ in the central region.
The D0 cuts were $|\eta| < 3.4$ and $p_t > 5$ GeV/$c$
in the central region.
In the forward region at the TeVatron, the request was for $2.5 < |\eta|
< 5.5$ and $p_t > 1.5$ GeV/$c$.  At the SSC, the central region was 
determined to be $ |\eta| < 2.5$ and $p_t > 10$ GeV/$c$, and the
 forward region given was $1.5 < |\eta| < 5.5$ and $p_t > 1.5$ GeV/$c$.
The calculations are done with cuts in rapidity not pseudorapidity,
 but the difference should
be small.  Table 3. shows the results for these cuts.

\vglue 0.25in
\centerline{Table 3. Cross Sections with Cuts Implemented.}
\begin{center}
\begin{tabular}{|c|c|c|c|c|}
\hline
CDF & D0 & TeVatron & SSC & SSC \\
Central & Central & Forward & Central & Forward \\
\hline
7.2 $\mu$b & 13 $\mu$b & 7.0 $\mu$b & 62 $\mu$b & 300 $\mu$b \\
\hline
\end{tabular}
\end{center}
The forward region results include the sum of the positive and negative 
rapidity results.  The result for 
the central SSC region seems low until one considers the large $p_t$-cut made.
Also, the large rapidity coverage of D0 helps considerably in enlarging
the cross section.

For additional enlightenment, we have plotted $d\sigma /dp_t$ versus $p_t$
for the central and forward regions for both the TeVatron and the SSC.  Before
we discuss the $d\sigma/dp_t$ plots we would also refer interested readers
to \cite{bm} for rapidity distributions giving additional useful information.
In Figure 1., we see that the expanded rapidity coverage of D0 makes the
cross section larger by a
factor of two over CDF rather uniformly over the entire $p_t$-range.  Most
of the cross section lies in the low-$p_t$ range.  Therefore if one could
lower the $p_t$-cut, the event increase would be sizeable.  For these plots,
we have chosen $M^2 = p_t^2 + m_{\rm b}^2$.  Also, these plots were produced
by running the programs for the Born approximation $p_t$-distributions and
multiplying by the 'K-factors' previously introduced; 2.2 for the TeVatron
plots and 3.2 for the SSC plots.  The justification for this was 1) time
was of the essence and the higher-order calculations would have taken 
a day each to compute and 2) in discussions \cite{js}, it was revealed that
the higher-order calculations generally raise the Born approximation results by
a fairly uniform amount across the entire $p_t$-range.
Figure 2. shows a dramatic fall-off in the forward region as $p_t$ increases,
again with most of the cross section in the low-$p_t$ region.  In the low-$p_t$
range, the cross section is reduced by a factor of three to five compared
to the central region. depending on the cut made.
Turning to the SSC, Figure 3. shows that by imposing a $p_t$-cut of 
10 GeV/$c$, most of the cross section is lost in the central region.
At large $p_t$, we find that the contribution is still appreciable.
Finally,  in the forward region, Figure 4. reveals the large-$p_t$
region is again still significant,
but again the majority of the cross section comes from the low-$p_t$ region.
The loss of cross section as $p_t$ increases is not so dramatic as it
is in the forward region at the TeVatron.

\mysection{Conclusions}

What can we conclude from these results?  First, the fixed-target results
are probably solid, since we can see from Figure 5. the results from UA1
 \cite{ua1}
are in good agreement
with the ${\cal O}(\alpha_S^3)$ results, and the energies for the fixed-target
experiments are lower than that of UA1.
Looking at Figure 6., we compare the ${\cal O}(\alpha_S^3)$ calculations of
$^1,^2$ with the 1988-89 and 1992-93 results of CDF \cite{cdf}.  Some of these
 data are still preliminary,
 of course, but it appears that the data do not fit
the calculation.
From the figure caption we see that we are off by a about a factor of 2.6.  But
we have some consolation because the shape is approximately correct, although
a slightly steeper distribution as discussed in \cite{bmq} would fit better.
  This 
factor of 2.6 will only be magnified when we look at the results for the SSC.
Clearly, we have a problem.

What are the possible solutions?  Calculate
the ${\cal O}(\alpha_S^4)$ corrections and see what difference that makes.  
That
is an enormous endeavor and would take years.  Try to make further headway
on the small-$x$ front.  This is possible but large uncertainties remain.
As an example, one interesting mechanism to accomodate the CDF data shown in
Figure 6. is to alter the form of the gluon distribution in the small-$x$
region \cite{bmq}.
But for a 'ballpark estimate' that probably is not too bad, why not do the
following:  try
\begin{eqnarray}
\sigma_{exp} = \sigma_0 e^{(K-1)}\,,
\end{eqnarray}
where $\sigma_0$ is the Born cross section, K is the appropriate 'K-factor,'
and $\sigma_{exp}$ is the expected cross section.
In the case of the TeVatron, $\sigma_0 = 17$ microbarns and $K = 2.2$.
We would get $\sigma_{exp} = 56$ microbarns.  For the SSC, $\sigma_0 = 170$
microbarns and $K = 3.2$.  Here $\sigma_{exp} = 1.5$ millibarns.  The 
distributions would also have the factor $e^{(K-1)}$ multiplying the
lowest-order distributions.  This is of course rather {\em ad hoc}, but the
results look reasonable.  More theoretically valid calculations are still
well off in the distance, and the numbers are needed now.

Finally, in the course of many discussions \cite{lst}, it was decided that
approximate cross section figures for each of the colliders, current and
proposed, should be provided so that an estimate of B-physics event
rates could be made.
Toward that end, we present Table 4., a compilation of cross section figures
that should be correct within a factor of two.
\vglue 0.25in
\centerline{Table 4. Cross Section Figures for Reference.}

\begin{center}
\begin{tabular}{|l|c|c|c|c|c|c|}
\hline
$\sqrt{S}$ & 43 GeV & 124 GeV & 200 GeV & 1.8 TeV & 15.4 TeV & 40 TeV \\
\hline
$\sigma$ & 20 nb  & 0.5 $\mu$b  & 2 $\mu$b  & 100 $\mu$b  & 0.5 mb & 1 mb\\
\hline
\end{tabular}
\end{center}

The numbers for the lower energies were arrived at essentially by rounding 
the results of the ${\cal O}(\alpha_S^3)$ calculation.  The 1.8 TeV result
was derived in the following way:  we took the fact that the curve that fits
the data of CDF is 2.6 times the ${\cal O}(\alpha_S^3)$ result.   
Multiplying the 37 microbarns by the factor of 2.6, we get a convenient
number of 100 microbarns for the TeVatron with no cuts.
The numbers for the LHC and the
SSC were based upon various estiamtes obtained using various parton
distribution sets.  They were also agreed upon in \cite{lst} and further 
detailed discussions about the uncertainties can be found in \cite{bm},
\cite{lst}.

Acknowledgements.
The authors would like to thank Jack Smith for his careful reading of the
manuscript.  This work was supported by the Lightner Sams Foundation, Inc. 
and the U.S. Department of Energy, Division of High Energy Physics, Contract
W-31-109-ENG-38.
\vfill
\newpage
%

\newpage
\begin{description}
\item[Fig. 1.]
$d\sigma /dp_t$ vs. $p_t$ for the kinematic cuts imposed
for the CDF collaboration (solid line) and the D0 collaboration (dashed line)
in the central region.
\item[Fig. 2.]
$d\sigma /dp_t$ vs. $p_t$ for the kinematic cuts imposed 
in the forward region at the TeVatron.
\item[Fig. 3.]
$d\sigma /dp_t$ vs. $p_t$ for the kinematic cuts imposed 
in the central region at the SSC.
\item[Fig. 4.]
$d\sigma /dp_t$ vs. $p_t$ for the kinematic cuts imposed 
in the forward region at the SSC.
\item[Fig. 5.]
$\sigma $ vs. $p_t^{\rm min}$ for $\sqrt{S} = 630$ GeV with $|y|
< 1.5$.  The data are taken from Table 2. of \cite{ua1}.  
The high curve was run
with  $m_{\rm b} = 4.5 $ GeV/$c^2$, and
$M = m_{\rm b}/2$.  The middle curve was run
with  $m_{\rm b} = 4.75 $ GeV/$c^2$, and
$M = m_{\rm b}$.  The low curve was run
with  $m_{\rm b} = 5.0 $ GeV/$c^2$, and
$M = 2m_{\rm b}$.  CTEQ1M distribution functions were used.
\item[Fig. 6.]
$\sigma $ vs. $p_t^{\rm min}$ for $\sqrt{S} = 1.8$ TeV with $|y|
< 1$.  The high solid curve was run
with $m_{\rm b} = 4.5 $ GeV/$c^2$, and
$M = m_{\rm b}/2$.  The middle solid curve was run
with  $m_{\rm b} = 4.75 $ GeV/$c^2$, and
$M = m_{\rm b}$.  The low solid curve was run
with  $m_{\rm b} = 5.0 $ GeV/$c^2$, and
$M = 2m_{\rm b}$.  CTEQ1M distribution functions were used.
The data with the thick error bars are taken from the 
88-89 and the thin error bars from the 92-93 runs of 
CDF \cite{cdf}.  The dashed curve is the
 middle solid curve multiplied by a factor of 2.6.
\end{description}
\end{document}